\documentclass[prl,twocolumn,superscriptaddress,showpacs,floatfix]{revtex4-1}
\usepackage{epsfig}
\usepackage{color}
\usepackage{graphics}
\begin{document}

\title{An orientationally ordered helical fluid phase in a quasi-one-dimensional system of confined hard spheres.}

\author{Mahdi Zaeifi Yamchi}
\affiliation{Department of Chemistry, University of Saskatchewan,
Saskatoon, Saskatchewan, S7N 5C9}

\author{Richard K. Bowles}\thanks{Corresponding Author: richard.bowles@usask.ca}
\affiliation{Department of Chemistry, University of Saskatchewan,
Saskatoon, Saskatchewan, S7N 5C9}
\email{richard.bowles@usask.ca}

\date{\today}

\begin{abstract}
We use a series of molecular dynamics simulations, and analytical theory, to demonstrate that a system of hard spheres confined to a narrow cylindrical channel exhibits a continuous phase transition from an isotropic fluid at low densities, to an orientationally ordered, but translationally disordered, helical fluid at high densities. The ordered fluid phase contains small sections of helix separated by topological defects that change the direction of the twist, altering the local chirailty. The defects break up the translational order, but the fluid develops long range orientational order. An analysis of the particle packings show that the length separation between defects controls the geometrical properties of the helical sections, including the orientation, and that pairs of defects experience a weak, but long range attraction resulting from entropic free volumes effects. These collective long range interactions overcome the restrictions on quasi-one-dimensional transitions, even though the particle-particle interaction is short ranged.
\end{abstract}

\maketitle

Spherical particles confined to quasi-one-dimensional structures, such as C$_{60}$ fullerenes packed into nanoscale tubes~\cite{Mickelson:2003kl}, or colloids trapped in self assembled channels~\cite{Jiang:2013dc}, form an array of complex helical particle arrangements. Hard spheres, which only interact through excluded volume, have also been shown close pack into a variety of single, double and triple, chiral helices, depending on the diameter of the confining channel~\cite{Erickson:1973tg,Pickett:2000p6160,Mughal:2011cs,Chan:2011bp,Mughal:2012jd}. Little is known about how these helical structures actually form. Arguments by van Hove~\cite{vanHove:1950kx,Lieb:1966wf} and Landau~\cite{Landau:1980tu} appear to rule out the possibility phase transition in quasi-one-dimensional systems with short ranged interactions because the entropic advantage of introducing a defect or domain wall always outweighs the energetic cost in the thermodynamic limit. However, a recent experiment by Lohr et al~\cite{Lohr:2010fa} suggested that a system of quasi-one-dimensional soft spheres confined in a narrow channel may exhibit a transition from a disordered fluid and an orientationally ordered fluid, that was similar in nature to the hexatic phase observed in bulk two-dimensional hard discs~\cite{Kosterlitz:1973uj,Nelson:1979iv,Young:1979hf,vonGrunberg:2007tr}.
Unfortunately, experimental restrictions prevented the necessary finite system size studies needed to demonstrate the thermodynamic nature of the transition.

\begin{figure}
\includegraphics[width=3.5in]{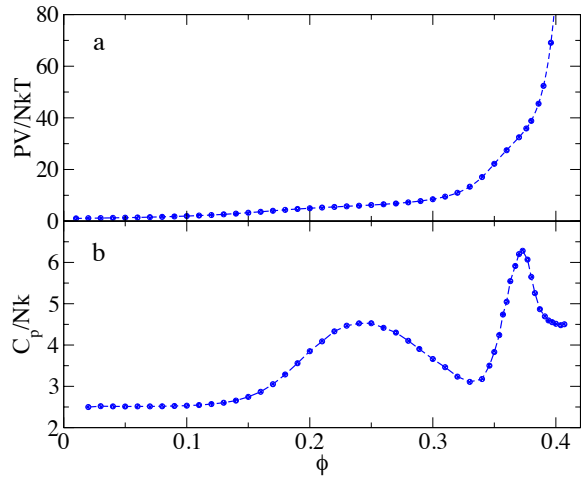}
\caption{(a) Equilibrium equation of state, $PV/NkT$ as a function of volume fraction, $\phi$ and (b), the isobaric heat capacity, $C_p/Nk$ as a function of $\phi$.}
\label{fig:eos}
\end{figure}

In this letter, we use molecular dynamics (MD) simulations to show that a continuous high order phase transition does exist and then use a geometric analysis of the packing to explore how this quasi-one-dimensional system is able to develop the necessary long range correlations and interactions.
We begin by using event driven MD simulations to determine the equilibrium properties of a system of $N=10000$ hard spheres of diameter $\sigma$, confined to a hard wall, cylindrical channel, with a diameter $H_d/\sigma=1.95$. At each volume fraction, $\phi=(2N\sigma^3)/(3H_dL)$, where $L$ is the length of the cylinder, $\left(200-10^{6}\right)N$ collisions were used to reach equilibrium and $\left(400-10^{7}\right)N$ collisions were used to collect data, depending on $\phi$. The system was compressed to the next $\phi$ using a modified version of the Lubachevsky and Stillinger~\cite{LUBACHEVSKY:1990p2185} (LS) algorithm that ensures $H_{d}/\sigma$ remains constant as the diameter of the spheres is changed ($L$ fixed) with a compression rate of $d\sigma/dt = 0.001$. The equation of state for the equilibrium fluid (EOS) (Fig.~\ref{fig:eos}a) varies continuously from the ideal gas limit, exhibiting a faint shoulder near $\phi=0.25$ and another near $\phi=0.35$, before it diverges at a $\phi$ just below the jamming density,$\phi_J$, of the perfect helix. Quasi-one-dimensional systems usually exhibit a single peak in the isothermal heat capacity, $C_p/Nk$, associated with the continuous translational ordering of the particles as defects are removed, but the appearance of a second maxima (Fig.~\ref{fig:eos}b) suggests the possibility of an additional ordering process.

\begin{figure}
\includegraphics[width=3.5in]{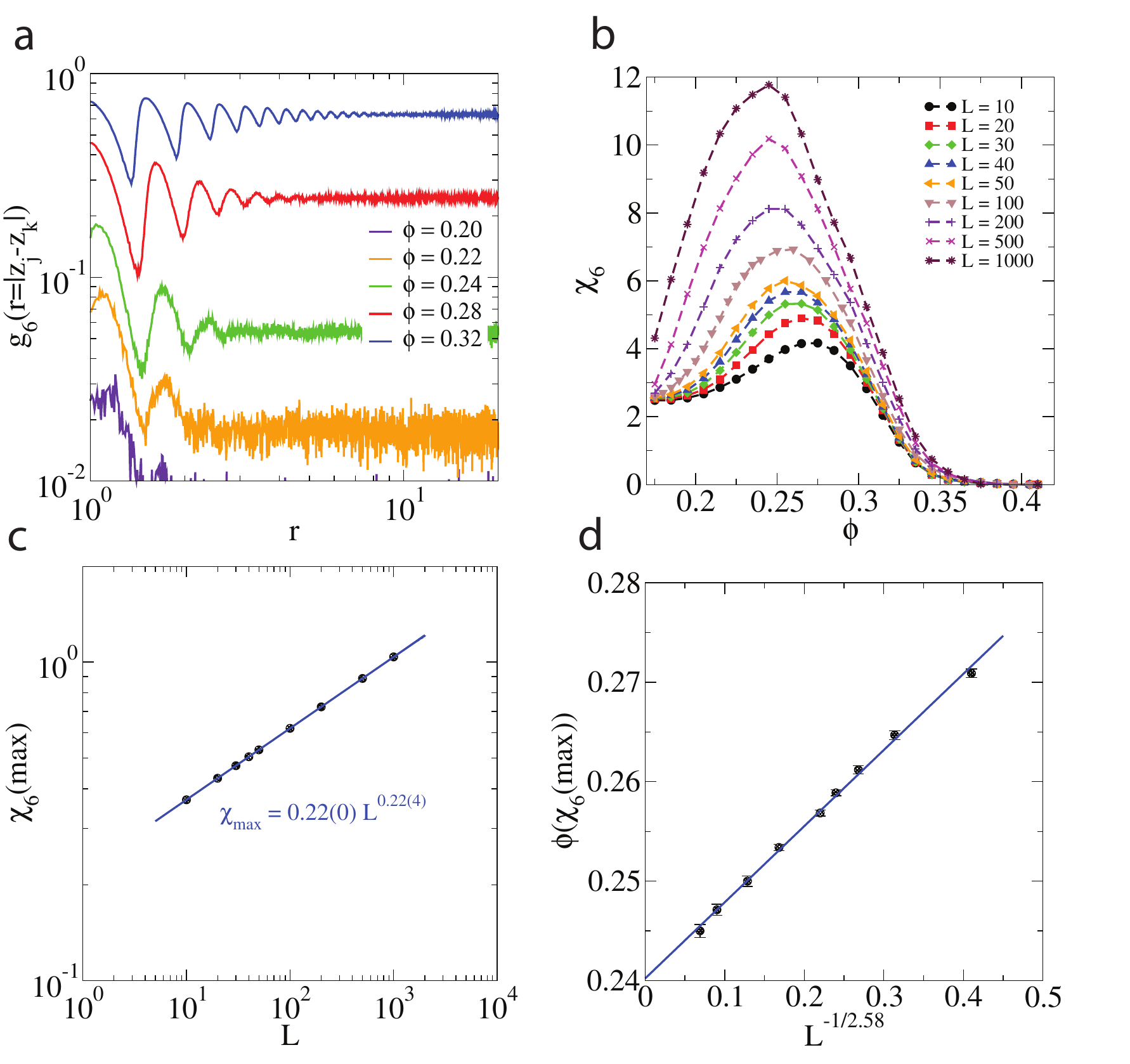}
\caption{(a) $g_6$ as a function of $r=z_k-z_j$, at different $\phi$. (b) $\chi_6$ as a function of $\phi$ for different system lengths $L$. (c) System size scaling for $\chi_6(\max)$. The data points are obtained from the maximum in (b) and the solid line represent the best fit to the data $\chi_6(\max)=0.22(0)L^{0.22(4)}$. (d) The volume fraction of the maximum in $\chi_6$, $\phi(\chi_6(\max))$,  as a function $L^{1/2.58(3)}$. The solid line represents the best fit to the data, $\phi_6(\max)=0.24(0)+0.07(6)L^{1/2.58(3)}$.}
\label{fig:orient}
\end{figure}

To characterize the local orientational order around particle $j$, we calculate $\psi_{6j}=e^{6i\gamma_{j}}$, where $\gamma_{j}$ is the in--plane angle formed between the first neighbours ($j-1,j+1$) along the channel. This is closely related to the parameter used by Lohr~\cite{Lohr:2010fa} and is coupled with the helical ordering. The average $\psi_6=\left<\psi_{6j}\right>$ is near zero at $\phi=0.2$ for larger system sizes, but increases with increasing densities before plateauing as it approaches unity in the high density limit (see figure in supplementary information, Fig S1). The orientational correlations along the channel, measured by $g_6(r=z_j-z_k)=\left<\psi_6j^*\psi_6k\right>$,  reach a constant value at high $\phi$, suggesting the existence of long range correlated orientational order (Fig~\ref{fig:orient}a). At lower $\phi$, $g_6$ begins to decay and we would expect long range orientational order to be non--existent in the disordered ideal gas limit. A finite system size analysis of the orientational susceptibility, $\chi_6=L(\left<|\psi_6|^2\right>-\left<|\psi_6|\right>^2)$ is carried out by considering  sub-systems of length $L$, in units of $\sigma$. Figure~\ref{fig:orient}b shows that $\chi_6$ exhibits a peak that increases in size with increasing $L$. Furthermore, the peak height at the maximum, $\chi_6(\max)$, scales as a power law, with an exponent $\nu=0.22(4)$ over two orders of magnitude in $L$ (Fig~\ref{fig:orient}c). The peak position also approaches its thermodynamic value as a power law and locates the transition at $\phi=0.24$ (Fig~\ref{fig:orient}d), which coincides with the location of the low density peak in the $C_p$.

\begin{figure}
\includegraphics[width=3.5in]{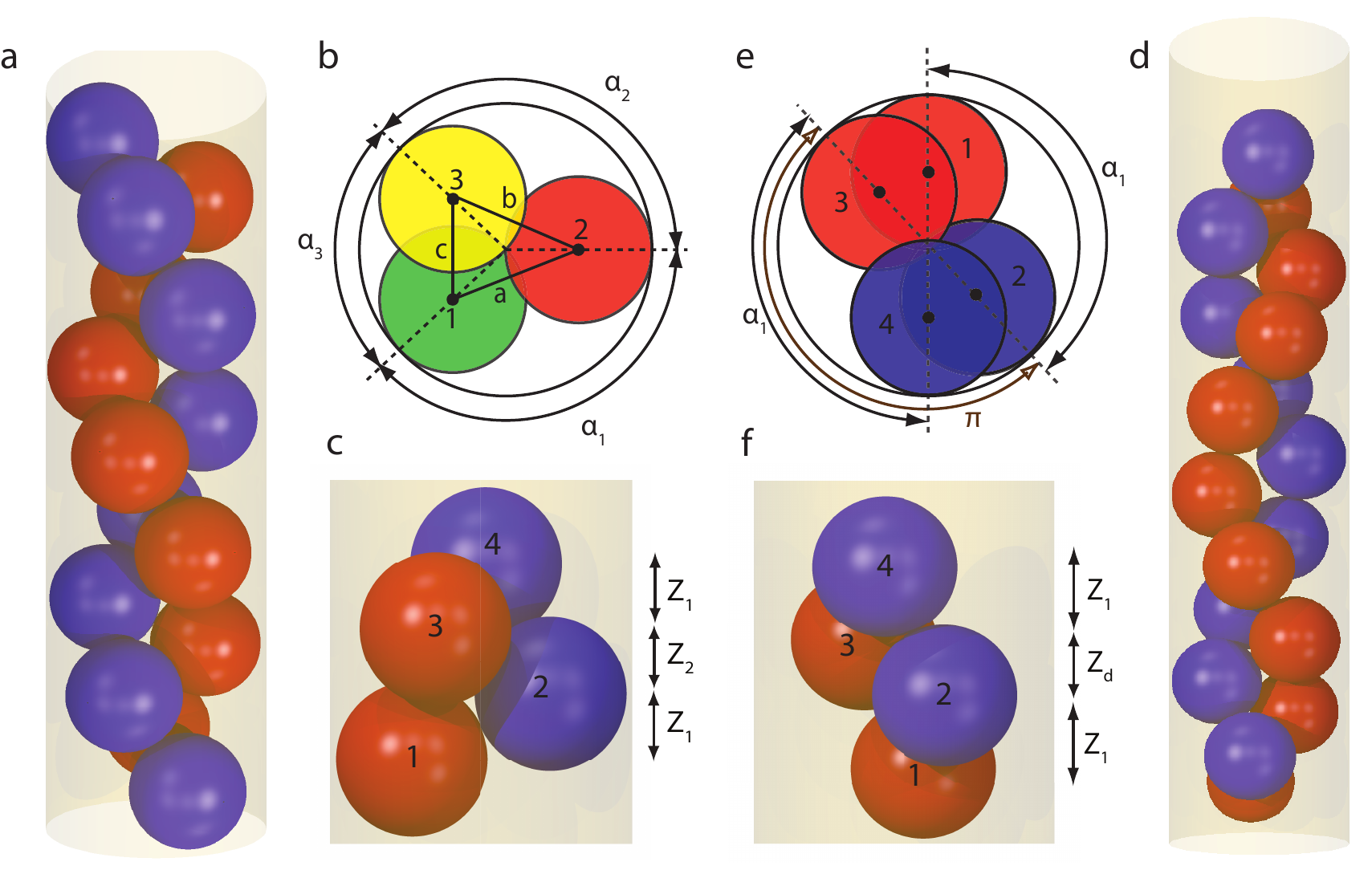}
\caption{(a) Perfect helical packing with $H_d/\sigma=1.95$. Alternating particles along the channel are coloured red and blue to highlight the direction of the helical twist. (b) Top view identifying the helical angles $\alpha_1$ and $\alpha_2$ between particles in a section helix. (c) Side view identifying the longitudinal lengths, $Z_1,Z_2$, in a segment of the helix. (d) A single defect causes the direction of the helical twist to change. (e) Top view of the four particles involved in a defect. Particles 2 and 3 are not in contact and are separated by an angle $\pi$. (f) Side view of the defect identifying $Z_d$.}
\label{fig:geo}
\end{figure}

\begin{figure}
\includegraphics[width=3.3in]{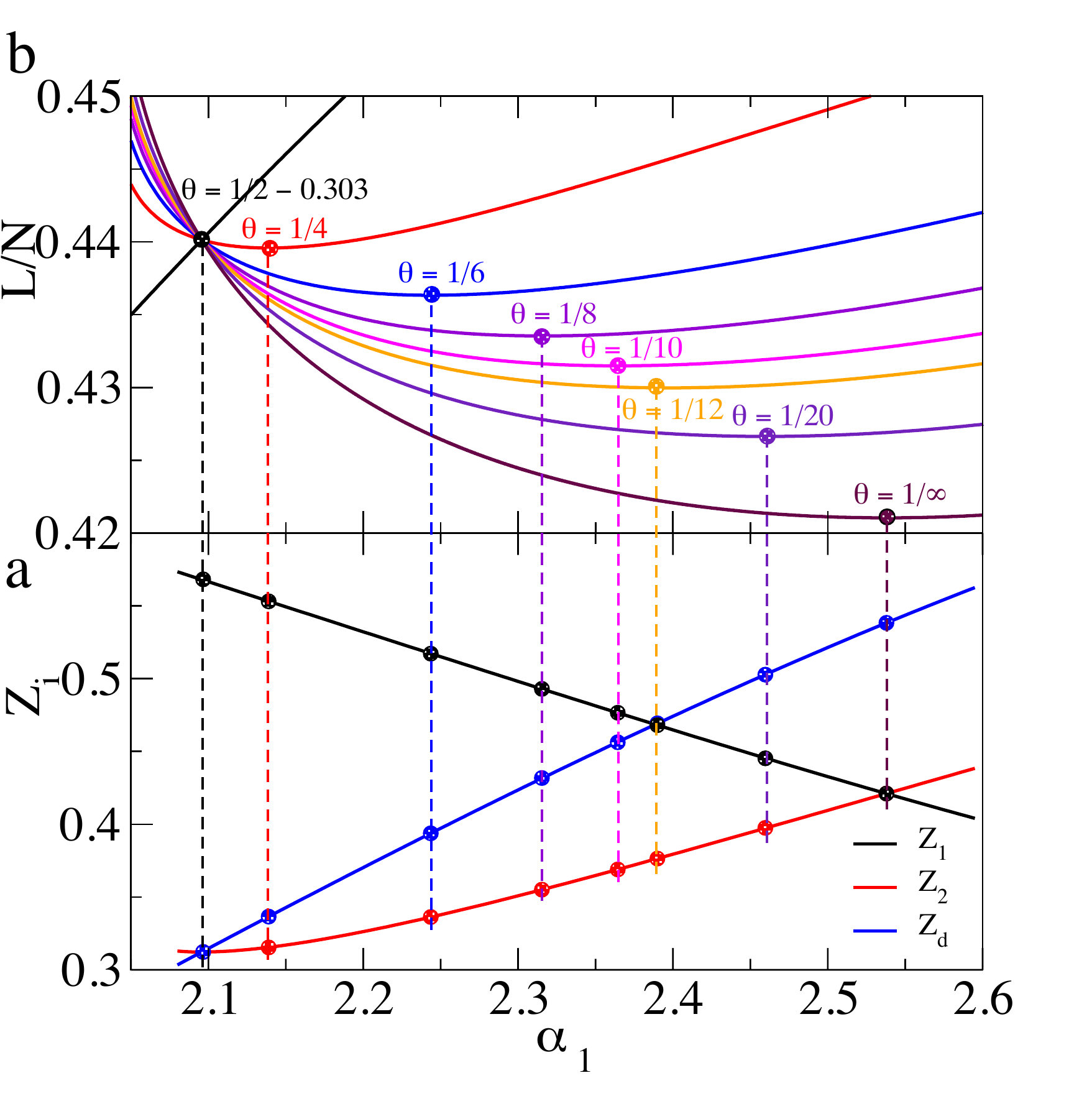}
\caption{(a) The longitudinal particle separations, $Z_1,Z_2$ and $Z_d$  as a function of $\alpha_1$ for the analytical model (solid lines) and simulations of defects crystals with a given $\theta$ (solid points). (b) $L/N$ as a function of $\alpha_1$ for the defect crystals with different $\theta$. The minimum represents the collectively jammed configuration and the values of $\alpha_1$ at the minima match the values of the simulated defect crystals. $\theta=0$ represents the prefect helix. The dotted lines connect the minima points with their geometric properties $Z_i$.}
\label{fig:cryst}
\end{figure}

To understand how this system appears to violate the restrictions on quasi-one-dimensional phase transitions, we examine the nature of the jammed packings accessible to the systems in an inherent structure analysis~\cite{Stillinger:1964jp,Stillinger:1982jb,Ashwin:2013eb,Yamchi:2012vg}, using both simulation and analytical theory. In three dimensions, a particle must have at least four rigid contacts, not all in the same hemisphere, to be locally jammed. With $H_d/\sigma$ in the range $1+\sqrt{3}/2 < H_d/\sigma < 1+4\sqrt{3}/7$, a particle can have up to five contacts, four sphere-sphere contacts with its two nearest neighbours on either side in the channel and one contact with the wall. The perfect helix in this system (Fig~\ref{fig:geo}a) can be constructed by considering the geometry associated with a particle in contact with the wall and its first and second neighbours (Fig~\ref{fig:geo},b,c). With particle one fixed, particle two can be placed in contact with the wall and the first particle, at a cylindrical angle $\alpha_{1}$ and a longitudinal displacement $Z_1$. Particle three can be placed at angle $\alpha_2$ and displacement $Z_2$. To continue to build the perfect helix, subsequent particles are placed at alternating increments of $\alpha_1,Z_1$ and $\alpha_2,Z_2$. Figure~\ref{fig:cryst}a shows how $Z_1$ and $Z_2$ varies as a function of $\alpha_1$ for the locally jammed structure (see supplemental for more details of the geometry).  However, while all the particles satisfy the local jamming condition, the system is not collectively jammed as $\alpha_1$ can be freely varied. To find the most dense jammed structure it is necessary to minimize the  length per particle, $L/N=0.5(Z_1+Z_2)$. This occurs when $Z_1=Z_2$, and correspondingly, $\alpha_1 = \alpha_2 = 2.5378$, forming a symmetrical packing with a jamming density $\phi_J=(2N\sigma^3)/(3H_dL) = 0.4164$, which is consistent with earlier studies~\cite{Pickett:2000p6160,Mughal:2011cs,Chan:2011bp,Mughal:2012jd}.

Jammed packings with a lower density can be formed by introducing pairs of topological defects, that change the chirality of sections of the helix (Fig.~\ref{fig:geo}d,e,f). With particles one and two fixed as before, the third particle is placed at an angle $\pi$ and a distance $Z_d$. Particles two and three do not actually contact and they only become locally jammed once the fourth particle is added with an angle $\alpha_1$, rotating in the opposite direction so that it establishes a contact with particle 2. Figure~\ref{fig:cryst}a shows how $Z_d$ varies as a function of $\alpha_1$ under the contact constraints. The sections of helix between defects are constructed in the same way as the perfect helix. As the defects simply replace the $Z_2$ length in the perfect structure we can write $L/N=0.5(Z_1+Z_2)-\theta(Z_d-Z_2)$, where $\theta$ is the concentration of defects and we assume the defects are regularly spaced, forming defect ``crystals". This ensures $\alpha_1$ is the same on both sides of the defect. The value of $\alpha_1$ that minimizes the $L/N$ now changes as a function of the defect concentration and the structure of the entire helix is altered (Fig~\ref{fig:cryst}b). The helical regions between defects are now asymmetrical with $Z_1\neq Z_2$ and $\alpha_1\neq\alpha_2$. 

The mechanical stability of the packings is studied by constructing defect crystal packings, with different defect concentrations over the full range $\theta=0-0.5$, near their predicted jamming density, $\phi_J$.  We then perform MD simulations that decompress and then re-compress the system in increments of $\Delta\phi=10^{-4}$, using $10^{6}N$ and $10^{7}N$ collisions to reach equilibrium and collect data, respectively. All the packings follow the free volume equation of state, $p = PV/NkT = 1/\delta = d/(1-\phi/\phi_{J})$ with a pressure that diverges at the predicted value (See supplemental information: Fig. S5). When the system is compressed back to jamming, the particle geometries are the same as those obtained from the model construction, corresponding to the minima points in fig.~\ref{fig:cryst}, and the defects have remained in their original locations, which suggest the packings are mechanically stable. To examine the inherent structures sampled by the fluid, we equilibrate the system at a volume fraction $\phi=0.01$, then compress a series of independent starting configurations to their jamming point using different compression rates (See supplementary information for more details: Fig S6). The resulting packings are polycrystalline in nature, consisting of individual subsections of helix,  each characterized by a particular $Z_1,\alpha_1$ and $Z_2,\alpha_2$, separated from other helical subsections by a defect that changes the chirality. These random packings are also found to be stable and follow the free volume equation of state when decompressed and re-compressed. Similar polycrystalline packings were observed by Lohr~\cite{Lohr:2010fa}.

\begin{figure}
\includegraphics[width=3.3in]{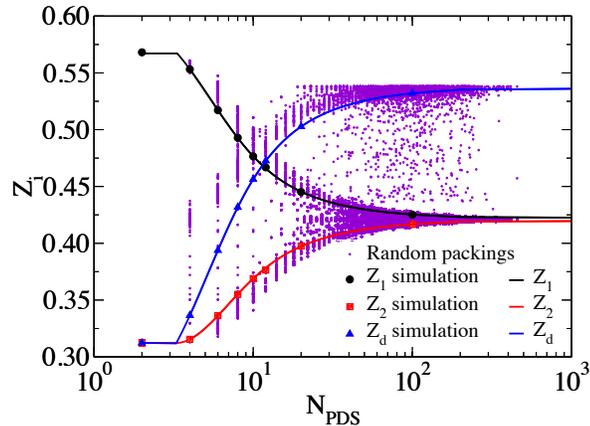}
\caption{$Z_i (i=1,2,d)$ as a function of defect separation, $N_{PDS}$ for the defect crystal model (minimum values of $L/N$ in Fig~\ref{fig:cryst}) (solid lines), simulations of the defect crystals (large symbols) and random packings formed through compression of the fluid (scatter points).}
\label{fig:random}
\end{figure}

Figure~\ref{fig:random} shows a scatter plot of the longitudinal $Z_i$ particle separations for the random packings as a function of the defect separation, measured in terms of the number of particles, $N_{PDS}$. They follow the general trends of the defect crystal model, but the $Z_1$ and $Z_d$ values for the random packings are higher than the defect crystals, while the $Z_2$ values are lower. Some of the scatter in the random packing results stem from the difficulty in forming exactly jammed structures during compression. However, the polycrystalline nature of the packing also means the defects connect subsections of helix with different geometric properties, i.e. with different values of $Z_1,Z_2$, which may also influence the packings. The appearance of odd numbered defect separations, especially at larger values of $N_{PDS}$, suggest the possibility of new local packing arrangements not considered in the defect crystal model. Nevertheless, Figure~\ref{fig:random} shows that the geometric properties of the subsections of the helix are determined by the $N_{PDS}$ in both the random packings and the defect crystals.
\begin{figure}
\includegraphics[width=3.5in]{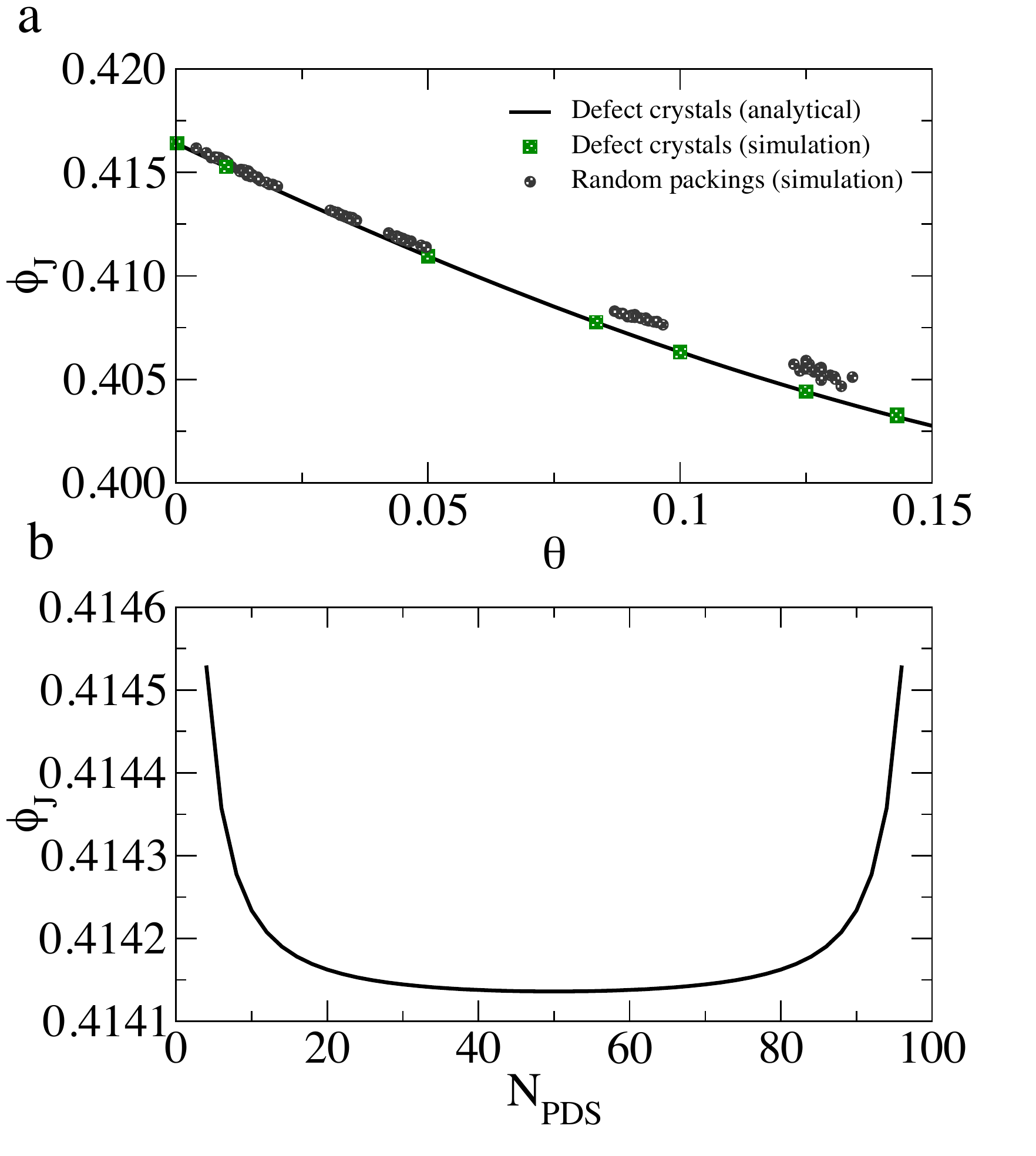}
\caption{(a) $\phi_J$ as a function of $\theta$ showing the random packings have a higher $\phi_J$ than the defect crystals. (b) $\phi_J$ as a function of $N_{PDS}$ for a model system containing two defects.}
\label{fig:interact}
\end{figure}

The random packings achieve a higher $\phi_J$ than the defect crystal packings for a given $\theta$ (Fig.~\ref{fig:interact}a) by pushing defects together, to form large sections of helix with a larger $N_{PDS}$ and a higher local density, at the cost of having a few low density regions. This shows that the distribution of the defects within the structure, and not just the number of defects, influences the jamming density. It also suggests there is a packing mediated interaction between defects that can be estimated by calculating the $\phi_J$ of the packing as a function of defect separation for a system containing just two defects. 
Figure~\ref{fig:interact}b shows $\phi_J$ for a system of $N=100$ particles, with periodic boundaries, and containing two defects. The first defect is placed at the origin. The properties of the two sections of helix and the defects are given by the defect crystal model with $\theta=1/N_{PDS}$ and $\theta=1/(N-N_{PDS})$ respectively. $\phi_J$ increases as the two defects move closer together, which induces a weak, but long range attraction between the defects because the fluid can increase its vibrational entropy by sampling deeper basins on the inherent structure landscape.  As the defects are pushed together, the helical angles in the subsection of helix become correlated over longer distances. When $N_{PDS}$ is larger than $N/2$, the defect effectively approaches another defect, which again causes $\phi_J$ to increase and highlights the long range nature of the interaction. This is the same entropic driving force~\cite{Frenkel:1999wj} that causes the bulk hard sphere system to freeze~\cite{Alder:1957vu}. However, the configurational entropic advantage of having defects in the quasi-one-dimensional system means that it is still favourable for the fluid to retain the defects, leading to the appearance of the polycrystalline random packings and preventing the formation of a solid phase. As a result, the system forms an orientationally ordered fluid containing topological defects that break up the translational order. The defect concentration in the equilibrium fluid phase can be measured directly by identifying where the local twist of the helical conformation changes direction. $\theta$ in the fluid decreases slowly in the low density phase, but it plateaus at the phase transition density. The defects are eventually eliminated in a rapid but continuous way at high densities, associated with the second $C_p$ maximum, as the system approaches close packing (see supplementary information for more details and Fig. S4).

The fluid phase described in this work is similar in nature to the hexatic phase observed in the bulk two dimensional hard disc model, where the topological defects are responsible for breaking up the long range translational order. Furthermore, the fundamental nature of the excluded volume interaction, which gives rise to the free volume entropic effects, suggests it is likely that this new phase behaviour could arise in a wide class of helical, quasi-one-dimensional systems. 

\begin{acknowledgements} 
We would like to thank C. Soteros, W.--K. Qi, P. H. Poole, I. Saika-Voivod and J. Whitehead for discussions and comments. Funding was provided by NSERC. Computational resources were provided by WestGrid and Compute Canada.
\end{acknowledgements}

%


\end{document}